\documentclass[aps,twocolumn,showpacs]{revtex4}
\usepackage{epsfig}
\usepackage{amsbsy}
\usepackage{amsmath}
\usepackage{amsfonts}
\usepackage{graphicx}
\usepackage{latexsym}

\begin{document}

\title{Entanglement in spin-one Heisenberg chains}
\author{XiaoGuang Wang$^1$, HaiBin Li$^2$, Zhe Sun$^1$, and You-Quan Li$^1$}
\affiliation{1, Zhejiang Institute of Modern Physics, Department
of Physics, Zhejiang University, HangZhou 310027, China}
\affiliation{2, Department of Applied Physics, Zhejiang University
of technology, HangZhou 310014, China}
\date{\today}
\begin{abstract}
By using the concept of negativity, we study entanglement in
spin-one Heisenberg chains. Both the bilinear chain and the
bilinear-biquadratic chain are considered. Due to the SU(2)
symmetry, the negativity can be determined by two correlators,
which greatly facilitate the study of entanglement properties.
Analytical results of negativity are obtained in the bilinear
model up to four spins and the two-spin bilinear-biquadratic
model, and numerical results of negativity are presented. We
determine the threshold temperature before which the thermal state
is doomed to be entangled.
\end{abstract}
 \maketitle

\section{Introduction}
Since Haldane predicts that the one-dimensional Heisenberg chain
has a spin gap for integer spins~\cite{Haldane}, the physics of
quantum spin chains has been the subject of many theoretical and
experimental studies. In these studies, the bilinear spin-one
Heisenberg model and the bilinear-biquadratic Heisenberg model
have played important roles~\cite{Affleck,Millet,Xiang}. The
corresponding Hamiltonians are given by
\begin{align}\label{Hamiltonian}
H_1=&\sum_{i=1}^N J{\bf S}_i\cdot {\bf S}_{i+1},\\
H_2=&\sum_{i=1}^N \left[J{\bf S}_i\cdot {\bf S}_{i+1}+\gamma({\bf
S}_i\cdot {\bf S}_{i+1})^2\right],
\end{align}
respectively. Here, we have assumed the periodic boundary
condition, and obviously, these two Hamiltonians exhibits a SU(2)
symmetry. Moreover, the bilinear-biquadratic model exhibits very
rich phase diagram~\cite{Schollwock}.

Recently, the study of entanglement properties in Heisenberg
systems have received much
attention~\cite{M_Nielsen}-\cite{QPT_GVidal}. Quantum entanglement
lies at the heart of quantum mechanics, and can be exploited to
accomplish some physical tasks such as quantum
teleportation~\cite{Tele}. Spin-half systems have been considered
in most of these studies. However, due to the lack of entanglement
measure for higher spin systems, the entanglement in higher spin
systems have been less studied. There are several proceeding works
on entanglement in spin-one chains. Fan et al.~\cite{Fan} and
Verstraete et al.~\cite{Verstraete} studied entanglement in the
bilinear-biquadratic model with a special value of $\gamma=1/3$,
i.e., the AKLT model~\cite{Affleck}. Zhou et al. studied
entanglement in the Hamiltonian $H_2$ for the case of two
spins~\cite{Zhou}.

In this paper, by using the concept of negativity~\cite{Vidal}, we
study pairwise entanglement in both the bilinear and the
bilinear-biquadratic Heisenberg spin-one models. For the case of
higher spins, a non-entangled state has necessarily a positive
partial transpose (PPT) according to the Peres-Horodecki
criterion~\cite{PH}. In the case of two spin halves, and the case
of (1/2,1) mixed spins, a PPT is also sufficient. However, in the
case of two spin-one particles, a PPT is not sufficient.
Nevertheless, the negative partial transpose (NPT) gives a
sufficient condition for entanglement, and due to the SU(2)
symmetry in the systems, the NPT is expected to fully capture the
entanglement properties.

The Peres-Horodecki criterion give a qualitative way for judging
if the state is entangled. The quantitative version of the
criterion was developed by Vidal and Werner~\cite{Vidal}. They
presented a measure of entanglement called negativity that can be
computed efficiently, and the negativity does not increase under
local manupulations of the system. The negativity of a state
$\rho$ is defined as
\begin{equation}
{\cal N(\rho)}=\sum_i|\mu_i|,
\end{equation}
where $\mu_i$ is the negative eigenvalue of $\rho^{T_2}$, and $T_2$ denotes \\
the partial transpose with respect to the second system. The
negativity ${\cal N}$ is related to the trace norm of $\rho^{T_2}$
via
\begin{equation} {\cal N(\rho)}=\frac{\|\rho^{T_2}\|_1-1}{2},
\end{equation}
where the trace norm of $\rho^{T_2}$ is equal to the sum of the
absolute values of the eigenvalues of $\rho^{T_2}$. If ${\cal
N}>0$, then the two-spin state is entangled.

We study entanglement in both the ground state and the thermal
state. The state of a system at thermal equilibrium described by
the density operator $\rho(T)=\exp(-\beta H)/Z$, where
$\beta=1/k_BT$, $k_B$ is the Boltzmann's constant, which is assume
to be 1 throughout the paper , and $Z=\text{Tr}\{\exp(-\beta H)\}$
is the partition function. The entanglement in the thermal state
is called thermal entanglement.

We organize the paper as follows. In Sec.~II, we give the exact
forms of the negativity for an SU(2)-invariant state, and show how
the negativity is related to two correlators. We also give that
how to obtain negativity from the ground-state energy and
partition function in the bilinear-biquadratic model. We study
entanglement in the bilinear and bilinear-biquadratic models in
Sec.~III and IV, respectively. Some analytical and numerical
results of negativity are obtained. We conclude in Sec. V.

\section{Negativity and correlators}
Schliemann considered the entanglement of two spin-one particles
via the Peres-Horodecki criteria~\cite{Schliemann}, and find that
the SU(2)-invariant two-spin state is entangled if either of the
following inequalities holds
\begin{align}
\langle({\bf S}_i\cdot {\bf S}_{j})^2\rangle>2, \nonumber\\
\langle({\bf S}_i\cdot {\bf S}_{j})^2\rangle+\langle {\bf
S}_i\cdot {\bf S}_j\rangle<1.
\end{align}
Now we explicitly give the expression of negativity for the
SU(2)-invariant two-spin state.

According to the SU(2)-invariant symmetry, any state of two
spin-one particles have the general form~\cite{Schliemann}
\begin{align}
\rho=&G|S=0,S_z=0\rangle\langle S=0,S_z=0|\nonumber\\
&+\frac{H}3\sum_{S_z=-1}^1|S=1,S_z\rangle\langle
S=1,S_z|\nonumber\\
&+\frac{1-G-H}5\sum_{Sz=-2}^2|S=2,S_z\rangle\langle S=2,S_z|,
\end{align}
where $|S,S_z\rangle$ denotes a state of total spin $S$ and $z$
component $S_z$, and
\begin{align}
G=&\frac{1}3[\langle({\bf S}_i\cdot {\bf
S}_{j})^2\rangle-1],\nonumber\\
H=&1-\frac{1}2[\langle{\bf S}_i\cdot {\bf
S}_{j}\rangle+\langle({\bf S}_i\cdot {\bf S}_{j})^2\rangle].
\end{align}

In order to perform partial transpose, the product basis spanned
by $\{|S_1=1,S_{1z}\rangle\otimes|S_2=1,S_{2z}\rangle\}$ is a
natural choice. By using the Clebsch-Gordan coefficients, we may
write state $\rho$ in the product basis. The partially transposed
with respect to the second spin $\rho^{T_2}$ can be written in a
block-diagonal form with two $1\times 1$ block, two $2\times 2$
block, and one $3\times 3$ block. After diagonalization of each
block, one find that the following only two eigenvalues of
$\rho^{T_2}$ are possibly negative~\cite{Schliemann},
\begin{align}
\mu_1=&\frac{1}6(2-\langle({\bf S}_i\cdot {\bf
S}_{j})^2\rangle),\nonumber\\
\mu_2=&\frac{1}3(\langle{\bf S}_i\cdot {\bf
S}_{j}\rangle+\langle({\bf S}_i\cdot {\bf S}_{j})^2\rangle-1).
\end{align}
Moreover, $\mu_1$ and $\mu_2$ occur with multiplicities 3 and 1,
respectively. Therefore, the negativity is obtained as
\begin{align}\label{neg}
{\cal N}^{(ij)}=&\frac{1}2\max[0,\langle({\bf S}_i\cdot{\bf
S}_j)^2\rangle-2]\nonumber\\
+&\frac{1}3\max[0,1-\langle({\bf S}_i\cdot{\bf
S}_j)\rangle-\langle({\bf S}_i\cdot{\bf S}_j)^2\rangle].
\end{align}
We see that for the SU(2)-invariant state, the negativity is
completely determined by two correlators $\langle({\bf
S}_i\cdot{\bf S}_j)\rangle$ and $\langle({\bf S}_i\cdot{\bf
S}_j)^2\rangle$.

Recall that the swap operator between two spin-one particles is
given by
\begin{equation}
{\cal S}_{ij}={\bf S}_i\cdot{\bf S}_j+({\bf S}_i\cdot{\bf
S}_j)^2-I
\end{equation}
where $I$ denotes the $9\times 9$ identity matrix. Then, the
negativity can be written in the following form
\begin{align}\label{negnewform}
{\cal N}^{(ij)}=&\frac{1}2\max[0,\langle{\cal
S}_{ij}\rangle-\langle{\bf S}_i\cdot{\bf
S}_j\rangle-1]\nonumber\\
+&\frac{1}3\max[0,-\langle{\cal S}_{ij}\rangle].
\end{align}
We see that if the expectation value $\langle{\cal
S}_{ij}\rangle<0$, the state is entangled. The swap operator
satisfies ${\cal S}_{ij}=1$, and thus it has only two eigenvalues
$\pm 1$. If a state is a eigenstate of the swap operator, the
expression (\ref{negnewform}) can be simplified. When the
corresponding eigenvalue is 1, Equation (\ref{negnewform})
simplifies to
\begin{equation}\label{nnn1}
{\cal N}^{(ij)}=\frac{1}2\max[0,-\langle{\bf S}_i\cdot{\bf
S}_j\rangle],
\end{equation}
and when the eigenvalue is -1, the equation simplifies to
\begin{equation}\label{nnn2}
{\cal N}^{(ij)}=\frac{1}3+\frac{1}2\max[0,-\langle{\bf
S}_i\cdot{\bf S}_j\rangle-2].
\end{equation}
In the former case, the state is entangled if $\langle{\bf
S}_i\cdot{\bf S}_j\rangle<0$, and in the latter case, the state is
an entangled state, and the negativity is larger than or equal to
1/3.

Now we consider the bilinear-biquadratic spin-one Heisenberg model
described by the Hamiltonian $H_2$. By applying the
Hellmann-Feynman theorem to the ground state of $H_2$ and
considering the translational invariance, we may obtain the
correlators as
\begin{equation}\label{negcc1}
\langle({\bf S}_i\cdot{\bf
S}_{i+1})\rangle=\frac{1}N\frac{\partial E_\text{GS}}{\partial
J},\; \langle({\bf S}_i\cdot{\bf
S}_{i+1})^2\rangle=\frac{1}N\frac{\partial
E_\text{GS}}{\partial\gamma},
\end{equation}
where $E_\text{GS}$ is the ground-state energy. Substituting the
above equation into Eq.~(\ref{neg}) yields
\begin{align}\label{negg1}
{\cal N}^{(ii+1)}=&\frac{1}2\max\Big[0,\frac{1}N\frac{\partial
E_\text{GS}}{\partial\gamma}-2\Big]\nonumber\\
+&\frac{1}3\max\Big[0,1-\frac{1}N\frac{\partial
E_\text{GS}}{\partial J}-\frac{1}N\frac{\partial
E_\text{GS}}{\partial\gamma}\Big].
\end{align}
For the case of finite temperature, we have
\begin{align}\label{negg2}
{\cal N}^{(ii+1)}=&\frac{1}2\max\Big[0,\frac{-1}{N\beta Z}\frac{\partial Z}{\partial\gamma}-2\Big]\nonumber\\
+&\frac{1}3\max\Big[0,1+\frac{1}{N\beta Z}\frac{\partial
Z}{\partial J}+\frac{1}{N\beta Z}\frac{\partial
Z}{\partial\gamma}\Big].
\end{align}
We see that the knowledge of ground-state energy (partition
function) is sufficient to determine the negativity for the case
of zero temperature (finite temperature).

\section{Bilinear Heisenberg model}
Let us know consider the entanglement in the bilinear Heisenberg
model. Due to the nearest-neighbor character of the interaction,
the entanglement between two nearest-neighbor spins is prominent
compared with two non-nearest-neighbor spins. Thus, we focus on
the nearest-neighbor case in the following discussions of
entanglement.

\subsection{Two spins}
For systems with a few spin, we aim at obtaining analytical
results of negativity. The Hamiltonian for two spins can be
written as
\begin{equation}\label{H2}
H_1={\bf S}_1\cdot {\bf S}_2=\frac{1}{2}[({\bf S}_1+{\bf
S}_2)^2-{\bf S}_1^2-{\bf S}_2^2],
\end{equation}
from which all the eigenvalues of the system are given by
\begin{equation}\label{evalue}
E_0=-2 (1),\; E_1=-1 (3), \; E_2=1 (5),
\end{equation}
where the number in the bracket denotes the degeneracy.

We investigate the entanglement of all eigenstates of the system.
When an energy level of our system is non-degenerate, the
corresponding eigenstate is pure. When a $k$-th energy level is
degenerate, we assume that the corresponding state is an equal
mixture of all eigenstates with energy ${\cal E}_k$. Thus, the
state correspoding to the $k$-th level with degeneracy becomes a
mixed other than pure, keeping all symmetries of the Hamiltonian.
A degenerate ground state is called thermal ground state in the
sense that it can be obtained from the thermal state $\exp[-H/(k_B
T)]/Z$ by taking the zero-temperature limit~\cite{M_Osborne}. The
$k$-th eigenstate $\rho_k$ can be considered as the thermal ground
state of the nonlinear Hamiltonian $H^\prime$ given by
$H^\prime=(H-{E}_k)^2$. Note that Hamiltonian $H^\prime$ inherits
all symmetries of Hamiltonian $H$.

As we consider interaction of two spins, from Eqs.~(\ref{H2}) and
(\ref{neg}), we obtain another form of the negativity as
\begin{align}\label{NN2}
{\cal N}^{(12)}=&\frac{1}2\max[0,\langle H^2_1\rangle-2]\nonumber\\
&+\frac{1}3\max[0,1-\langle H_1+H^2_1\rangle].
\end{align}
To determine the negativity, it is sufficient to know the
cumulants $\langle H_1\rangle$ and $\langle H_1^2\rangle$.

From Eqs.~(\ref{NN2}) and (\ref{evalue}), the negativities
correspond to the $k$-th level are obtained as
\begin{equation}\label{Nvalue}
{\cal N}_0^{(12)}=1,\; {\cal N}_1^{(12)}=1/3, \; {\cal
N}_2^{(12)}=0.
\end{equation}
We see that the ground state is a maximally entangled state, the
first-excited state is also entangled, but the negativity of the
second-excited state is zero.

Having known negativities of all eigenstates, we next consider the
case of finite temperature. The cumulants can be obtained from the
partition function. From Eq.~(\ref{evalue}), the partition
function is given by
\begin{equation}
Z=e^{2\beta}+3e^{\beta}+5e^{-\beta}.
\end{equation}
A cumulant of arbitrary order can be calculated from the partition
function,
\begin{align}\label{cumu}
\langle
H_1^n\rangle=&\frac{(-1)^n}{Z}\frac{\partial^n}{\partial\beta^n}Z\nonumber\\
=&\frac{(-1)^n}{Z}\left[2^ne^{2\beta}+3e^\beta+5(-1)^ne^{-\beta}\right]
\end{align}
Substituting the cumulants with $n=1,2$ to Eq.~(\ref{NN2}) yields
\begin{align}\label{N1}
{\cal
N}=&\frac{1}{2Z}\max\left(0,2e^{2\beta}-3e^{\beta}-5e^{-\beta}\right)\nonumber\\
&+\frac{1}{3Z}\max\left(0,3e^\beta-e^{2\beta}-5e^{-\beta}\right).
\end{align}
Thus, we obtain the analytical expression of the negativity.

The second term in Eq.~(\ref{N1}) can be shown to be zero. To see
this fact, it is sufficient to show that $F(x)=x^3-3x^2+5>0$,
where $x=e^{\beta}>1$. It is direct to check that the function $F$
takes its minimum 1 at $x=2$. As the minimum is large than zero,
the function is positive definite. Thus, equation~(\ref{N1})
simplifies to
\begin{equation}\label{N}
{\cal
N}=\frac{1}{2Z}\max\left(0,2e^{2\beta}-3e^{\beta}-5e^{-\beta}\right).
\end{equation}
The behavior of the negativity versus temperature is similar to
that of the concurrence~\cite{Conc} in the spin-half Heisenberg
model~\cite{M_Arnesen}, namely, the negativity decreases as the
temperature increases, and there exists a threshold value of
temperature $T_\text{th}$, after which the negativity vanished.
This behavior is easy to understand as the increase of temperature
leads to the increase of probability of the excited states in the
the thermal state, and the excited states are less entangled in
comparison with the ground state. From Eq.~(\ref{N}), the
threshold temperature can be analytically obtained as
\begin{align}
T_{\text{th}}=&\frac{1}{\ln
(\frac{1}2+\frac{1}{2(11+2\sqrt{30})^{1/3}}+\frac{(11+2\sqrt{30})^{1/3}}{2})}\nonumber\\
\approx& 1.3667.
\end{align}

\subsection{Three spins}
The Hamiltonian for three spins is rewritten as
\begin{equation}\label{H3}
H_1=\frac{1}{2}[({\bf S}_1+{\bf S}_2+{\bf S}_3)^2-{\bf S}_1^2-{\bf
S}_2^2-{\bf S}_3^2],
\end{equation}
from which the ground-state energy and the correlator $\langle{\bf
S}_1\cdot{\bf S}_2\rangle$ are immediately obtained as
\begin{equation}\label{c31}
E_\text{GS}=-3,\;\langle{\bf S}_1\cdot{\bf S}_2\rangle=-1.
\end{equation}
In order to know the ground-state negativity, we need to
calculator another correlator $\langle({\bf S}_1\cdot{\bf
S}_2)^2\rangle$.

By considering the translational invariance and using similar
techniques given by Refs.\cite{Kouzoudis,Schnack,Lin}, the
ground-state vector is obtained as
\begin{align}\label{psi3}
|\Psi\rangle_{\text{GS}}=\frac{1}{\sqrt{6}}(&|012\rangle+|201\rangle+|120\rangle\nonumber\\
-&|021\rangle-|102\rangle-|210\rangle),
\end{align}
where $|n\rangle$ denote the state $|s=1,m=s-n\rangle$, the common
eigenstate of ${\bf S}^2$ and $S_z$. Then, we can check that
\begin{equation}\label{relation3}
{\bf S}_1\cdot{\bf
S}_2|\Psi\rangle_{\text{GS}}=-|\Psi\rangle_{\text{GS}}.
\end{equation}
Thus, the correlator $\langle({\bf S}_1\cdot{\bf S}_2)^2\rangle$
is found to be
\begin{equation}\label{c32}
\langle({\bf S}_1\cdot{\bf S}_2)^2\rangle=1.
\end{equation}
Substituting Eqs.~(\ref{c31}) and (\ref{c32}) to Eq.~(\ref{neg})
yields
\begin{equation}
{\cal N}=1/3.
\end{equation}
We see that spins 1 and 2 are in an entangled state at zero
temperature. With the increase of temperature, the negativity
monotonically decreases until it reaches the threshold value
$T_\text{th}=0.9085$, after which the negativity vanishes.

\subsection{Four spins}
Now we consider the four-spin case, and the corresponding
Hamiltonian can be written as
\begin{equation}\label{H3}
H_1=\frac{1}{2}[({\bf S}_1+{\bf S}_2+{\bf S}_3+{\bf S}_4)^2-({\bf
S}_1+{\bf S}_3)^2-({\bf S}_2+{\bf S}_4)^2].
\end{equation}
The standard angular momentum coupling theory directly yields the
ground-state energy and the correlator $\langle{\bf S}_1\cdot{\bf
S}_2\rangle$
\begin{equation}\label{c44}
E_\text{GS}=-6,\;\langle{\bf S}_1\cdot{\bf S}_2\rangle=-3/2.
\end{equation}
Then, we need to compute another correlator $\langle({\bf
S}_1\cdot{\bf S}_2)^2\rangle$ or alternatively the expectation
value $\langle{\cal S}_{12}\rangle$. So, it is necessary to know
the exact form of the ground state.

By using similar techniques given by
Refs.\cite{Kouzoudis,Schnack,Lin}, the ground-state vector is
obtained as
\begin{align}\label{psi4}
|\Psi\rangle_{\text{GS}}=&1/2|\psi_1\rangle-3/2|\psi_2\rangle+|\psi_3\rangle\nonumber\\
&-3/2|\psi_4\rangle+3/\sqrt{2}|\psi_5\rangle+|\psi_6\rangle.
\end{align}
where
\begin{align}
|\psi_1\rangle=&1/2(|0022\rangle+|2002\rangle+|2200\rangle+|0220\rangle),\nonumber\\
|\psi_2\rangle=&1/2(|0112\rangle+|2011\rangle+|1201\rangle+|1120\rangle),\nonumber\\
|\psi_3\rangle=&1/2(|0121\rangle+|1012\rangle+|2101\rangle+|1210\rangle),\nonumber\\
|\psi_4\rangle=&1/2(|0211\rangle+|1021\rangle+|1102\rangle+|2110\rangle),\nonumber\\
|\psi_5\rangle=&1/\sqrt{2}(|0202\rangle+|2020\rangle),\nonumber\\
|\psi_6\rangle=&|1111\rangle.
\end{align}
Then, from the explicit form of the ground state, after two-page
calculations, we obtain the expectation value of the swap operator
as
\begin{equation}\label{sss}
\langle{\cal S}_{12}\rangle=1/6.
\end{equation}
Substituting Eqs.~(\ref{c44}) and (\ref{sss}) to
Eq.~(\ref{negnewform}) leads to
\begin{equation}
{\cal N}=1/3.
\end{equation}
It is interesting to see that the ground-state negativity in the
four-qubit model is the same as that in the three-qubit model. The
threshold value can be found to be $T_\text{th}=1.3804$.

For $N\ge 5$, it is hard to obtain analytical results of
negativity. The behaviors of negativity are similar to those for
$N\le 4$, namely, with the increase of temperature, the negativity
decreases until it vanishes at threshold temperature
$T_{\text{th}}$. For instance, the threshold temperatures
$T_\text{th}\approx 0.95$ and $T_\text{th}\approx 1.21$ for five
and six spins, respectively. The negativity for two
nearest-neighbors spins is estimated as ${\cal N}=0.1240$ (${\cal
N}=0.2509$) for the case of five spins (six spins).

\section{Bilinear-Biquadratic Spin-One Heisenberg Chain}
\begin{figure}
\includegraphics[width=0.45\textwidth]{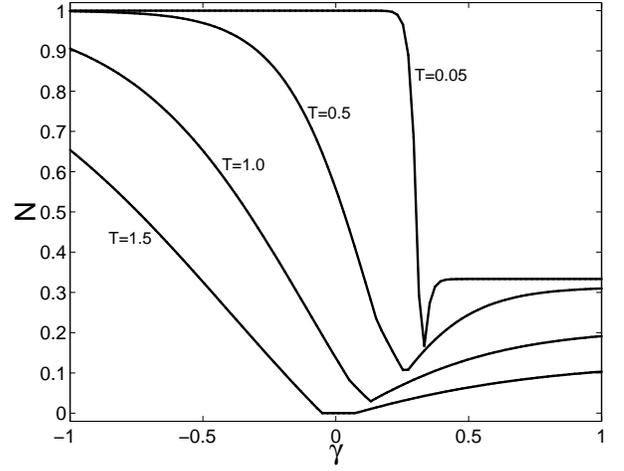}
\caption{Negativity versus $\gamma$ for different temperatures in
the two-spin model~($J=1$).}
\end{figure}
We now study entanglement properties in the bilinear-biquadratic
spin-one Heisenberg model, and first consider the case of two
spins. From Eq.~(\ref{negg1}) with $N=1$, if we know the
ground-state energy, the negativity is readily obtained. The
ground-state energy is given by
\begin{equation}\label{Egs}
E_\text{GS}=\left\{
\begin{array}{ll}
-2J+4\gamma& \;\text{when}\; \gamma<1/3, \\
-1J+\gamma & \;\text{when}\; \gamma>1/3,
\end{array}
\right.
\end{equation}
We see that there exits a level crossing at the point of
$\gamma=1/3$. Then, substituting the above equation into
Eq.~(\ref{negg1}) yields
\begin{equation}\label{Nbi}
{\cal N}=\left\{
\begin{array}{ll}
1& \;\text{when} \;\gamma<1/3, \\
1/3 & \;\text{when} \;\gamma>1/3,
\end{array}
\right.
\end{equation}
Before the point $\gamma=1/3$, the negativity of the ground-state
is 1, while the negativity of the first-excited state is 1/3.
After the cross point, the ground and first-excited interchanges,
and thus, the negativity of the ground state after the cross point
is 1/3. It is interesting to see that the model at the cross point
is just the AKLT model.

In Fig.~1, we plot the negativity versus $\gamma$ for different
temperatures. The level cross greatly affects the behaviors of the
negativity at finite temperatures. For a small temperature
($T=0.05$), the negativity displays a jump from 1 to 1/3 near the
cross point. For higher temperatures, the negativity first
decreases, and then increases at $\gamma$ increases from -1 to 1.
For $T=1.5$, we observe that there exists a range of $\gamma$, in
which the negativity is zero.

\begin{figure}
\includegraphics[width=0.45\textwidth]{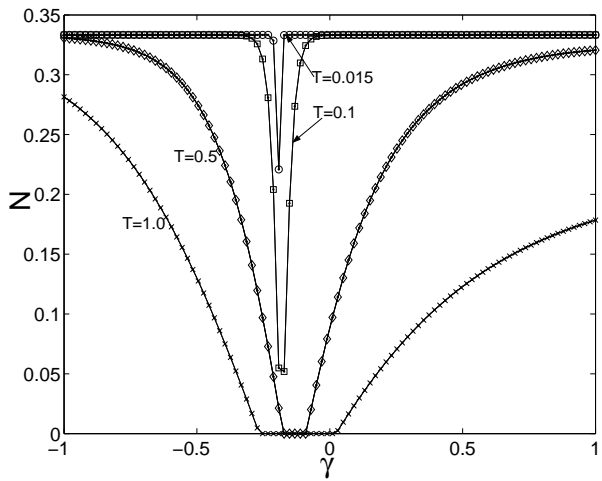}
\caption{Negativity versus $\gamma$ for different temperatures in
the three-spin  model ($J=1$).}
\end{figure}

\begin{figure}
\includegraphics[width=0.45\textwidth]{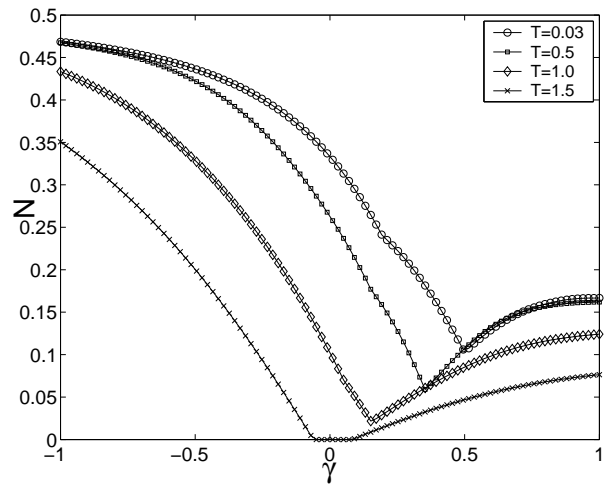}
\caption{Negativity versus $\gamma$ for different temperatures in
the four-spin  model ($J=1$).}
\end{figure}

For the three-spin case, we plot the negativity versus $\gamma$
for different temperatures in Fig.~2. For a low temperature
$T=0.015$, we observe a dip, which results from the level crossing
near the point of $\gamma=-0.2121$. When $T=0.1$, the dip becomes
more evident. For the cases of higher temperatures ($T=0.5$ and
$T=1.0$), there exists a range of parameter $\gamma$, in which the
negativity is zero.

For the four-spin case, we also a plot of the negativity for
different temperatures. For $T=0.03$, as $\gamma$ increases, the
negativity decreases until it reaches its minimum, and then
increases. For $T=0.5$ and $1.0$, the behaviors of negativity are
similar to the case of $T=0.01$, and the difference is that the
minima shifts left. There are some common features in the
behaviors of negativity for different number of spins. The maximum
value of negativity occurs at $\gamma=-1$; for higher
temperatures, there exists a range of $\gamma$, in which the
negativity is zero.

From Figs.~1-3, we observe that the thermal state is always
entangled at a lower temperature. When temperature increases, the
negativity decreases until it reaches zero, namely, the thermal
fluctuation suppresses entanglement. Before the threshold
temperature, the state is doomed to be entangled. We numerically
calculated the threshold temperature and the result are shown in
Fig.~4. The threshold temperature decreases nearly linearly when
$\gamma$ increases from -1 to a certain value of $\gamma$. After
reaching a minimum, it begin to increase. We see that the
behaviors of the threshold temperature are similar for different
number of spins.

As a final remark, we consider the following Hamiltonian
\begin{equation}
H_3=\sum_{i\neq j}^N J{\bf S}_i\cdot {\bf
S}_{j}=\frac{1}2\left(\sum_{i=1}^N {\bf S}_i\right)^2-N,
\end{equation}
where the interaction is between all spins, and there are all
together $N(N-1)/2$ terms. The system not only shows a SU(2)
symmetry, but also an exchange symmetry, namely, the Hamiltonian
in invariant under exchange operation ${\cal S}_{ij}H_3 {\cal
S}_{ij}=H_3$. For $N=2,3$, the model is identical to Hamiltonian
$H_2$. We know that the ground state is non-degenerate when
$N=2,3$, and thus it must be an eigenstate of $S_{ij}$ and
Eqs.~(\ref{nnn1}) and (\ref{nnn2}) can apply. From the angular
momentum coupling theory, the ground-state energy of $H_3$ is
readily obtained as $E_\text{GS}=-N$, and thus we have $\langle
{\bf S}_i\cdot {\bf S}_j\rangle=-2/(N-1)$. Then, from
Eqs.~(\ref{nnn1}) and (\ref{nnn2}), the negativity can be either
$1/(N-1)$ or $1/3$. For $N=2$ ($N=3$), the ground state is
symmetric (antisymmetric) and then the negativity is 1 (1/3),
consistent with previous results. However, for $N\ge 4$, the
ground-state is degenerate and we cannot apply Eqs.~(\ref{nnn1})
and (\ref{nnn2}). The numerical results show that the negativity
is zero for $N\ge 4$.

\begin{figure}
\includegraphics[width=0.45\textwidth]{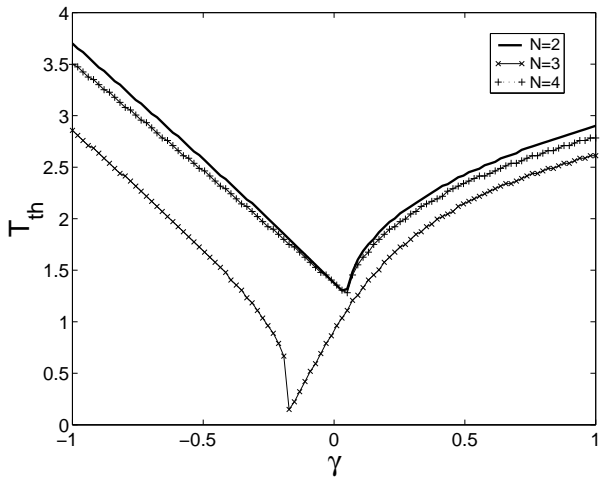}
\caption{Threshold temperature versus $\gamma$ for different
number of spins.}
\end{figure}

\section{Conclusions}
In conclusion, by using the concept of negativity, we have studied
entanglement in spin-one Heisenberg chains. Both the bilinear
model and bilinear-biquadratic model are considered. Although NPT
only give a sufficient condition for entanglement, due to the
SU(2) symmetry, we believe that this condition considerably
captures entanglement properties of the system. Moreover, the
negativity gives an upper bound to the distillation of
entanglement~\cite{Vidal}, one of the fundamental entanglement
measures. We have given explicitly the relation between the
negativity and two correlators. The merit of this relation is that
the two correlators completely determine the negativity and it
facilitates our discussions of entanglement properties.

We have obtained analytical results of negativity in the bilinear
model up to four spins and in the two-spin bilinear-biquadratic
model. We numerically calculated entanglement in the
bilinear-biquadratic model for $N=2,3,4$, and the threshold
temperatures versus $\gamma$ are also given.  We have restricted
us to the small-size systems, and aimed at obtaining analytical
results via symmetry considerations and getting some numerical
results via the exact diagonalization method. However, for larger
systems, the exact diagonalization method is not a viable route.
It is interesting to investigate large systems by some mature
numerical methods such as the quantum monte-carlo method and
density-matrix renomalization group method. And it is also
interesting to consider other SU(2)-invariant spin-one systems
such as the dimerized and frustrated systems.

\acknowledgements We thanks for the helpful discussions with G. M.
Zhang, C. P. Sun and Z. Song. This work is supported by the
National Natural Science Foundation of China under Grant No.
10405019.

\end{document}